\documentstyle[eqsecnum,floats,prd,aps,epsf]{revtex}

\draft
\begin{document}
\newcommand{\beq}{\begin{equation}}
\newcommand{\eeq}{\end{equation}}
\newcommand{\beqn}{\begin{eqnarray}}
\newcommand{\eeqn}{\end{eqnarray}}

\twocolumn[\hsize\textwidth\columnwidth\hsize\csname
@twocolumnfalse\endcsname

\title{Stability of coalescing binary stars against 
gravitational collapse: hydrodynamical simulations}

\author{Masaru Shibata}
\address{Department of Earth and Space Science,~Graduate School of 
	Science,~Osaka University,
	Toyonaka, Osaka 560-0043, Japan}
\author{Thomas W.~Baumgarte}
\address{Department of Physics, University of Illinois at Urbana-Champaign, 
	Urbana, Il 61801}
\author{Stuart L.~Shapiro}
\address{Departments of Physics and Astronomy, and NCSA, 
	University of Illinois at Urbana-Champaign, Urbana, Il 61801}
\maketitle

\begin{abstract}
We perform simulations of relativistic binary stars in post-Newtonian
gravity to investigate their dynamical stability prior to merger
against gravitational collapse in a tidal field. In general, our
equations are only strictly accurate to first post-Newtonian order,
but they recover full general relativity for spherical, static stars.
We study both corotational and irrotational binary configurations of
identical stars in circular orbits.  We adopt a soft, adiabatic
equation of state with $\Gamma = 1.4$, for which the onset of
instability occurs at a sufficiently small value of the compaction
$M/R$ that a post-Newtonian approximation is quite accurate. For such
a soft equation of state there is no innermost stable circular orbit,
so that we can study arbitrarily close binaries. This choice still
allows us to study all the qualitative features exhibited by any
adiabatic equation of state regarding stability against gravitational
collapse.  We demonstrate that, independent of the internal stellar
velocity profile, the tidal field from a binary companion stabilizes
a star against gravitational collapse.
\end{abstract}


\vskip2pc]

\section{Introduction}

Binary neutron stars are known to exist and for some of these systems
in our own galaxy (including PSR B1913+16 and B1534+12),
general relativistic effects in the binary orbit have been measured to
high precision~\cite{tamt93,tw89}. Interest in binary neutron stars has been
stimulated by the prospect of future observations of extragalactic
systems by gravitational wave interferometers 
like LIGO\cite{LIGO}, VIRGO\cite{VIRGO}, TAMA\cite{TAMA} and GEO\cite{GEO}. 
Binary neutron stars are among the most promising sources of
gravitational waves for these detectors, and therefore it is important
to predict theoretically the gravitational waveform emitted during the
inspiral and the final coalescence of the two stars. Interest in these
systems also arises on a more fundamental level, since the two-body
problem is one of the outstanding unsolved problems in classical
general relativity.

Considerable effort has gone into understanding binary neutron
stars. Most of this work has been performed within the framework of
Newtonian and post-Newtonian gravity (see, e.g.,~\cite{BCSSTc} for a
review and list of references). General relativistic treatments are
currently only in their infancy. Recently, Wilson, Mathews and
Marronetti~\cite{wilson} (hereafter WMM) reported results obtained
with a relativistic hydrodynamics code. Their code assumed several
simplifying physical and mathematical approximations.  Their results
suggest that the central densities of the stars increase as the stars
approach each other and that massive neutron stars, stable in
isolation, individually collapse to black holes prior to merger. WMM
therefore find that in general relativity, the presence of a companion
star and its tidal field tend to destabilize the stars in a binary
system.  This conclusion is contrary to what is expected from
Newtonian~\cite{lrs93}, post-Newtonian~\cite{lai,lrs97,wiseman},
perturbative~\cite{brady} and matched asymptotic
expansion~\cite{EEF,KIP} treatments of the problem. Constructing
self-consistent, fully relativistic initial data for two neutron stars
in a circular, quasi-equilibrium orbit does not show any evidence of
this ``crushing effect'' either~\cite{BCSSTa}. Moreover, applying
energy turning-point methods to sequences of these initial data
suggests that inspiraling neutron star binaries are {\em secularly}
stable all the way down to the innermost stable circular
orbit~\cite{BCSSTb}. To summarize, most researchers currently believe
that the maximum allowed rest mass of neutron stars in close binaries
is {\em larger} than in isolation, and that their central density is
{\em smaller} than in isolation. If there exists any destabilizing,
relativistic effect at high post-Newtonian order, then this effect is
much smaller than the dominating stabilizing effect of the tidal
field.

However, to date, the only fully {\em dynamical} treatment of the
problem in general relativity -- that of WMM -- reports a
star-crushing effect. In this paper, we perform a new, fully dynamical
simulation for binary stars in post-Newtonian gravity.  We use a
formalism in which (1) all first post-Newtonian terms are taken into
account, and (2) sufficient nonlinearity is retained, so that
spherical, static stars satisfy the fully general relativistic
equations exactly. As explained in section II below, this formalism
is very suitable for studying binary neutron stars. We study
relativistic effects in binary stars with $M/R \ll 1$, where 
$M$ and $R$ are typical values of the stellar mass and radius, 
so that a post-Newtonian treatment is completely adequate.

By performing a fully dynamical calculation, we can relax various
constraints assumed in previous treatments. For example, Wiseman
assumed the stars to remain spherically symmetric~\cite{wiseman},
Baumgarte {\em et al.}~\cite{BCSSTc} assumed the binary stars to be
corotating, and Thorne~\cite{KIP} assumed the stars' orbital
separation to be much larger than the stars' radius. Here, we relax
all these assumptions and study tidally deformed stars, both
corotational and irrotational, at arbitrarily small separations. We
still find that the presence of the tidal field of a companion star
tends to stabilize neutron stars against catastrophic collapse.

To establish the stability of binary stars against collapse, we
construct quasi-equilibrium initial data for identical binary neutron
stars in a close, circular orbit. The idea is to show whether stars in
a binary formed from the inspiral of objects which are stable in
isolation remain stable at close separation.  Our models have rest
masses near the maximum allowed rest mass for spherical stars in
isolation and thus provide the best candidates for collapse if the
tidal field is destabilizing (stars with rest masses well below the
maximum allowed value are unambiguously stable).  In order to
demonstrate that these stars are dynamically stable, we need to locate
the onset of instability in the binary, and compare it with the onset
of instability for isolated stars. Since the shift is fairly small, a
very careful treatment with high numerical accuracy is necessary. We
detail our method of locating the onset of instability in section IV.

The paper is organized as follows: In section II, we present the
post-Newtonian formalism adopted in this paper.  We calibrate our code
in section III by locating the analytically known onset of radial
instability of relativistic 
spherical stars against gravitational collapse~\cite{Chandra}. 
In section IV, we study the dynamical stability against gravitational
collapse of close binary stars, and briefly summarize our results in
section V.

\section{Formulation}

In the usual post-Newtonian treatment, the fluid and field equations
are derived by systematically expanding the Einstein equation and the
relativistic hydrodynamic equations in powers of
$c^{-1}$~\cite{ChandraPN}.  In this paper, we introduce a different
approach.  Since in a first order post-Newtonian approximation, the
spatial metric $\gamma_{ij}$ may always be chosen conformally flat, we
can derive post-Newtonian equations by starting with the
relativistic equations written previously in the conformally flat 
approximation~\cite{wilson,BCSSTc}. We then
neglect some of the  second and higher order post-Newtonian terms, 
but retain sufficient nonlinearity so
that this formalism recovers full general relativity for some limiting
regimes of interest in this paper.

We write the spatial metric in the form
\begin{equation}
\gamma_{ij}=\psi^4\tilde \gamma_{ij}=\psi^4 \delta_{ij}, 
\end{equation} 
so that the line element becomes
\beqn
ds^2&=&g_{\mu\nu}dx^{\mu}dx^{\nu} \nonumber \\
&=&(-\alpha^2+\beta_k\beta^k)dt^2
+2\beta_i dx^idt+\psi^4 \delta_{ij}dx^idx^j.
\eeqn
Here, $\alpha$, $\beta_i$ and $\psi$ are the lapse function, the shift
vector and the conformal factor and we adopt geometrized
units in which $c \equiv 1 \equiv G$.  We also adopt cartesian
coordinates $x^i=(x, y, z)$, so that the covariant derivative $\tilde
\nabla_k$ associated with $\tilde \gamma_{ij} = \delta_{ij}$
conveniently reduces to the ordinary partial derivative
$\partial/\partial x^k$.

We employ a perfect fluid stress-energy tensor
\beq
T^{\mu\nu}=\rho\left(1+\varepsilon+{P \over \rho}\right) u^{\mu}u^{\nu}
+Pg^{\mu\nu},
\eeq
where $\rho$, $\varepsilon$, $P$, and $u^{\mu}$ denote the 
rest mass density, specific internal energy, pressure, and the fluid
four velocity, respectively. For initial data, we assume a constant
entropy configuration with a polytropic pressure law 
\begin{equation}
P = K \rho^\Gamma,
\end{equation}
where $K$ and $\Gamma=1+1/n$ are constant and $n$ is the polytropic
index. For the evolution of the matter we assume the adiabatic
relation
\begin{equation}
P = (\Gamma-1)\rho\varepsilon.
\end{equation}

The continuity equation is 
\beq
{\partial \rho_* \over \partial t}
+{\partial (\rho_* v^i) \over \partial x^i}=0~,
\eeq
where $\rho_*=\rho \alpha u^0 \psi^6$ and $v^i=u^i/u^0$. 

The relativistic Euler equation is
\beqn
{\partial (\rho_* \tilde u_i) \over \partial t}
+{\partial (\rho_* \tilde u_i v^j) \over \partial x^j} 
&=&-\alpha \psi^6 P_{,i}-\rho_* \alpha \tilde u^0 \alpha_{,i} \nonumber \\
&& ~ +\rho_* \tilde u_j \beta^{j}_{~,i}
+{2\rho_* \tilde u_k \tilde u_k \over \psi^5 \tilde u^0} \psi_{,i}, 
\eeqn
and the energy equation is
\begin{equation}
{\partial e_* \over \partial t}+{\partial (e_* v^j) \over \partial x^j} = 0,
\end{equation}
where 
\beqn
&&\tilde u_j=(1+\Gamma \varepsilon)u_j ,\\
&&\tilde u^0=(1+\Gamma \varepsilon)u^0, \\
&&e_*=(\rho_* \varepsilon_*)^{1/\Gamma}; 
\varepsilon_*=\varepsilon (\alpha u^0 \psi^6)^{\Gamma-1}, \\
&&v^i=-\beta^i+{u_j \over \psi^4 u^0}. 
\eeqn
Note that $\beta^i=\psi^{-4}\beta_i$ in the conformal flat 
approximation. From the normalization condition $u^i u_i = - 1$, we find
\beqn
(\alpha u^0)^2& &=1+{u_i u_i \over \psi^4} \nonumber \\
& &=1+{\tilde u_i \tilde u_i \over \psi^4}\biggl[
1+\Gamma{\varepsilon_* \over (\alpha u^0 \psi^6)^{\Gamma-1} }\biggr]^{-2}. 
\eeqn
In our numerical simulation, we use $\rho_*$, $\tilde u_i$, and $e_*$
as the independent variables that are determined by the
hydrodynamical equations.

Equations for $\psi$, $\beta^i$ and $\alpha$ can be found from 
the Hamiltonian constraint, the momentum constraint, and the maximal
slicing condition tr$K = 0$, where tr$K$ is the trace of the 
extrinsic curvature $K_{ij}$ (details can be found, for example, 
in~\cite{BCSSTc}).

Assuming maximal slicing for all times, we have $\partial_t \mbox{tr}K
= 0$, and can use the trace of the evolution equation for $K_{ij}$ to find
\beq
\Delta (\alpha \psi)=2\pi\alpha \psi^5
\Bigl(E + 2S_{ij}\delta^{ij}\psi^{-4}\Bigr)
+{7 \over 8}\alpha \psi^5 K_{ij} K^{ij}.
\eeq
Here $E=\rho(1+\Gamma \epsilon)(\alpha u^0)^2-P$, 
$S_{ij}=T_{ij}$, and $\Delta$ is the flat space Laplacian. 

Since we assume the spatial metric to remain conformally flat for all 
times, the trace free part of the time evolution equation for 
$\gamma_{ij}$ has to vanish, which yields
\beq
2\alpha \psi^{-4}K_{ij} = \delta_{il}\beta^l_{~,j}+
\delta_{jl}\beta^l_{~,i}-{2 \over 3} \delta_{ij}\beta^l_{~,l}. 
\eeq
This equation shows that the extrinsic curvature no longer represents
independent dynamical degrees of freedom (i.e., it may no longer
exactly satisfy its fully relativistic evolution equation).  Inserting
this into the momentum constraint, $ (\psi^6 K^i_{~j})_{,i} = 8\pi J_j
\psi^6 $, where $J_i=\rho_* \tilde u_i \psi^{-6}$, we find
\beqn
\Delta \beta^i+{1 \over 3}\beta^j_{~,ji}&=&
\biggl[ \ln\Bigl({\alpha \over \psi^{6}} \Bigr) \biggr]_{,j} 
\Bigl(\beta^i_{~,j}+\beta^j_{~,i}-{2 \over 3}\delta_{ij}\beta^l_{~,l}\Bigr)
\nonumber \\
&+&16\pi\alpha J_i. 
\eeqn

Finally, the Hamiltonian constraint yields 
\beq
\Delta \psi=-2\pi \psi^5E-{K_{ij}K^{ij} \psi^5 \over 8}.
\eeq

For the post-Newtonian point of view assumed here, a conformally flat
spatial metric takes into account all Newtonian and first
post-Newtonian terms, and differences from a general, fully nonlinear
metric appear at second post-Newtonian order~\cite{GS,ASF}. We
can therefore use the above equations, which assume a conformally flat
metric, as a starting point for a post-Newtonian approximation.  We
simplify the problem by neglecting other terms of second or
higher post-Newtonian order. In particular, we will neglect the
nonlinear terms $7\alpha\psi^5K_{ij}K^{ij}/8$,~ $[\ln(\alpha
\psi^{-6})]_{,j}(\beta^i_{~,j}+\beta^j_{~,i} -2 \delta_{ij}
\beta^l_{~,l}/3)$, and $K_{ij}K^{ij}\psi^5/8$. Note that for static,
spherically symmetric spacetimes, these terms vanish identically,
so that we still recover full general relativity for these
spacetimes.

Adopting this approximation, the field equations reduce to
\beqn \label{ell1}
&&\Delta (\alpha \psi)=2\pi\alpha \psi^5
\Bigl(E + 2S_{ij}\delta^{ij}\psi^{-4} \Bigr)\equiv 4\pi S_{\alpha\psi} ,\\
&&\Delta \beta^i+{1 \over 3}\beta^j_{~,ji}=16\pi\alpha J_i, \\
&& \Delta \psi=-2\pi \psi^5E \equiv 4\pi S_{\psi}.
\eeqn
We decompose the equation for $\beta^i$  using 
\beq
\beta^i=4B_i-{1 \over 2}\biggl[ \chi_{,i}+(B_k x^k)_{,i}\biggr]
\eeq
so that $B_i$ and $\chi$ satisfy
\beqn
&&\Delta B_i=4\pi\alpha J_i,\\
&&\Delta \chi=-4\pi\alpha J_i x^i \label{ell2}. 
\eeqn

To summarize, we have reduced Einstein's equations to six elliptic
equations for the six functions $\alpha\psi$, $\psi$, $B_i$ and
$\chi$. We solve these equations together with the boundary conditions
\beqn
&&\alpha\psi=1 - {1 \over r}\int S_{\alpha\psi} dV +O(r^{-3}),\\
&&\psi = 1 - {1 \over r}\int S_{\psi} dV +O(r^{-3}),\\
&&B_x=-{x \over r^3}\int \alpha J_x xdV-{y \over r^3}\int \alpha J_x ydV
\nonumber \\  &&\hskip 2cm +O(r^{-4}),\\
&&B_y=-{x \over r^3}\int \alpha J_y xdV-{y \over r^3}\int \alpha J_y ydV
\nonumber \\ &&\hskip 2cm +O(r^{-4}),\\
&&B_z=-{z \over r^3}\int \alpha J_z zdV+O(r^{-4}),\\
&&\chi={1 \over r}\int \alpha J_i x^i dV + O(r^{-3}),
\eeqn 
where $dV$ is the coordinate volume element.
Note that having removed some of the nonlinear terms from the elliptic
equations~(\ref{ell1}) to~(\ref{ell2}), their right hand sides now
have compact support. This further simplifies the computations, since
in imposing the boundary conditions at a finite separation, we do 
not truncate any source terms in~(\ref{ell1}) to~(\ref{ell2}) that
extend to infinity. 

Having neglected some second post-Newtonian terms, our formalism is
strictly only first-order post-Newtonian. Note, however, that we have
only truncated some of the non-linear terms in the field equations,
pieces of which, loosely speaking, can be associated with dynamical
features of the gravitational fields. In particular, we still solve
the fully relativistic hydrodynamic equations.  We therefore retain
many of the nonlinear features of full general relativity, and expect
that this formalism provides an excellent approximation in 
several limiting regimes of interest here.

For example, for static, spherically symmetric stars, we recover the
fully relativistic, Oppenheimer-Volkov solution. This is, because we
can choose coordinates such that all the terms neglected in the field
equations vanish identically, and the equations of hydrodynamics (or,
in this case, hydrostatics) are fully relativistic. Constructing
sequences of equilibrium solutions, we can apply the energy turning
point method and find the onset of radial instability -- again without
approximation.

Corotational or irrotational binary stars at large separations are very
close to being spherically symmetric because 
the two stars interact only through weak tidal fields 
(for example, in Newtonian case, see refs.\cite{lrs93,lrs94}). 
Hence, our formalism can describe the individual stars to high accuracy, 
whether or not the stars are very compact and have strong gravitational
fields. We therefore expect that our approximations are quite adequate. 

In a binary in which the orbital separation $a$ is large compared to
the stellar radius $R$, we can treat the gravitational effects and
tidal deformation due to the companion star as a small perturbation.
We can then expand the gravitational field around the unperturbed,
spherical solution~\cite{EEF,KIP}
\beqn
\alpha & = & {}_{(0)}\alpha + _{(2)}\alpha \epsilon^2 + _{(4)}\alpha \epsilon^4
	\nonumber \\ 
	& & + _{(5)}\alpha \epsilon^5 + _{(6)}\alpha \epsilon^6
	+ _{(7)}\alpha \epsilon^7 + \cdot\cdot\cdot,\label{MAA}\\
\psi & = & {}_{(0)}\psi + _{(2)}\psi \epsilon^2 + 
	_{(4)}\psi \epsilon^4\nonumber \\
	& & + _{(5)}\psi \epsilon^5 + _{(6)}\psi \epsilon^6
	+ _{(7)}\psi \epsilon^7 + \cdot\cdot\cdot,\\
\beta^i & = & _{(1)}\beta^i \epsilon + _{(3)}\beta^i \epsilon^3
	+ _{(5)}\beta^i \epsilon^5 \nonumber \\
	& & + _{(6)}\beta^i \epsilon^6 + _{(7)}\beta^i\epsilon^7
 	+ _{(8)}\beta^i\epsilon^8 + \cdot\cdot\cdot,\\
\tilde \gamma_{ij} & = & \delta_{ij} + _{(2)}h_{ij} \epsilon^2 +
	_{(4)}h_{ij} \epsilon^4 + _{(5)}h_{ij}\epsilon^5 +\cdot\cdot\cdot,
\label{MAE}
\eeqn
where we assume 
\begin{equation}
v^i \sim u_i \sim (M/a)^{1/2} 
\alt (R/a)^{1/2} \sim \epsilon, 
\end{equation}
and where $M$ is gravitational mass, and $_{(0)}\alpha$ and
$_{(0)}\psi$ denote the spherical symmetric
solutions\cite{foot0}. Note that $\epsilon$ denotes the magnitude of
gravitational effects from the companion star, and hence the expansion
in terms of $\epsilon$ is different from a post-Newtonian expansion
(see~\cite{EEF}.)

In our formalism, we can calculate $_{(0)}\alpha$ and $_{(0)}\psi$
exactly. Newtonian and first post-Newtonian terms appearing in
$_{(n)}\alpha$, $_{(n)}\psi$ and $_{(n)}\beta^i$ for $n \not=0$ are
also taken into account consistently for all order in $\epsilon$,
although second post-Newtonian terms in these and $_{(n)}h_{ij}$ are
not included.  Our approximation is therefore appropriate for
investigating the stability of fully general relativistic spherical
stars due to Newtonian and first post-Newtonian tidal effects, 
which is our goal in this paper.

Details of our numerical methods, for solving both the hydrodynamical
equations and the Poisson equations, can be found in~\cite{SON}.  We
assume symmetry with respect to equatorial plane, and solve the
equations on a uniform grid of size $(2N+1, 2N+1,N+1)$, covering
the physical space $-L \leq x,y \leq L$ and $0 \leq z \leq L$ where
$L$ is location of outer boundaries. We use $N=50$ for the spherical
symmetric stars, and $N=50$, 60, and 75 for binary configurations.

As a numerical check, we monitor the conservation of proper mass
\beq
M_p=\int \rho_* dV,
\eeq
total gravitational mass
\beq
M \equiv - 2 \int S_{\psi} dV,
\eeq
and total angular momentum
\beq
J \equiv \int(-yJ_x+xJ_y)\psi^6 dV. 
\eeq
Our difference scheme guarantees conservation of $M_p$ exactly.
Accurate conservation of $M$ and $J$ depends on $N$, and for $N=75$,
the error in one orbital period is $\alt 0.05\%$ for $M$ and $\alt
0.5\%$ for $J$, respectively.

\section{Dynamical stability of spherical static stars}

In this section we calibrate our code by locating the analytically
known onset of instability of spherical equilibrium stars.

For initial conditions, we construct sequences of equilibria
satisfying the Oppenheimer-Volkov equations. Note that for these
configurations our formalism is exact and recovers the fully
relativistic solutions. In Fig.~1 we show the proper mass $M_p$ and
the (isotropic) coordinate radius $R$ as a function of the central
density $\rho_c$ for a polytrope with $\Gamma=1.4~(n=2.5)$. We have
taken advantage of the scale freedom in the problem and chosen $K =
1$. Together with $c = 1 = G$, this assignment uniquely determines our
non-dimensional units for length, mass, etc..  For any other value of
$K$, all the results can be rescaled trivially (see,
e.g.,~\cite{BCSSTa}).  Note that for a critical central density
$\rho_c=\rho_{\rm crit}(\simeq 1.2\times 10^{-4})$, the mass $M_p$
goes through a maximum $M_{\rm max} \sim 1.248$. This maximum marks
the onset of radial instability, and separates the stable branch
($\rho_c <
\rho_{\rm crit}$) from the unstable branch ($\rho_c > \rho_{\rm
crit}$) of the sequence.

For $\Gamma=1.4$, the compaction of the maximum mass configuration,
$M_p/R|_{\rm crit} \sim 0.03$, is much less than unity (recall
$M_p/R|_{\rm crit} \rightarrow 0$ as $\Gamma \rightarrow 4/3$).
Nevertheless, the mass versus central density equilibrium curve still
exhibits an extremum, and therefore has all the qualitative features
of any value of $\Gamma > 4/3$ regarding the issue of radial
stability. Choosing $\Gamma=1.4$ therefore allows us to study these
qualitative features in a regime in which the post-Newtonian
approximation is very reliable.

\begin{figure}[t]
\epsfxsize=2.5in
\begin{center}
\leavevmode
\epsffile{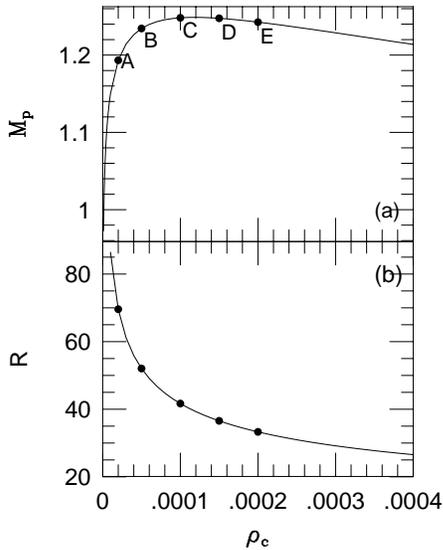}
\end{center}
\caption{Proper mass $M_p$ (top panel) and isotropic radius 
$R$ (botton panel) as a function of central density $\rho_c$ for 
relativistic, spherical polytopes with  $\Gamma=1.4$.}
\end{figure}

\begin{figure}[t]
\epsfxsize=2.5in
\begin{center}
\leavevmode
\epsffile{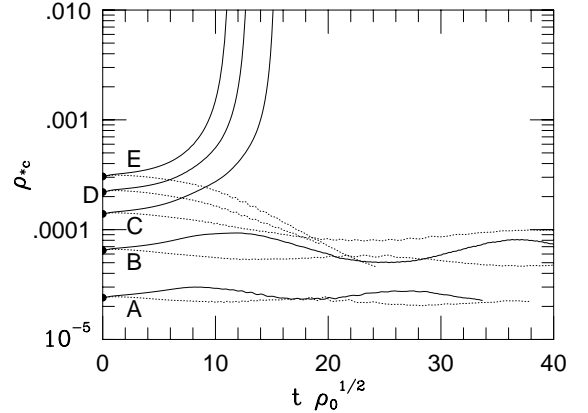}
\end{center}
\caption{Time evolution of $\rho_*$ at center of stars 
for models A -- E.}
\end{figure}

Consider results for five different initial configurations, which we
denote by A, B, C, D and E (see Fig.~1).  We construct these initial
data with a one-dimensional integration of the Oppenheimer-Volkov
equations.  These data are in equilibrium according to the
one-dimensional finite difference equations.  Interpolating these data
onto the three-dimensional grid of our evolution scheme introduces a
slight perturbation of the equilibrium solution, since the truncation
error of the three-dimensional finite difference equations is
different from the one-dimensional one.  Typically, we find that the
pressure of the interpolated models is slightly larger than the
required equilibrium value on the three-dimensional grid. We 
compensate for that by artificially decreasing the polytropic constant
$K$ by a small amount. Note that our finite differencing is
convergent, so that for finer grids we need to change $K$ by smaller
amounts. For the results presented here, the initial radius of the
star is covered by 30 grid points.

Note that because of the scale freedom in the problem, reducing $K$ is
equivalent to increasing the mass. This is, because $K^{n/2}$, where
$n$ is the polytropic index, has units of mass (or length) 
in geometrized units. Consider now a model
in which we have reduced $K$ by, say, a small fraction $\delta \ll
1$.  We can then rescale this model in order to investigate a
different $K$, for example, the original value $K = 1$.  
Then, the mass of the re-scaled
model has to increase by a fraction $n\delta/2$ to remain in
equilibrium.  For example, reducing $K$ by 0.5 \% is, for $n=2.5$,
equivalent to increasing the model's mass by 0.625 \%.

We can now test the stability of our models by varying $K$ by small
amounts. Small variations in $K$ serve to trigger small initial
perturbations away from equilibrium (e.g.~``pressure depletion'').
Stable models will not change their qualitative behavior,
whereas unstable models will. In Fig.~2 we show the time evolution of
the central density $\rho_*$ for the five models. Solid curves are the
results for $K=0.995$, and dotted curves are for $K=1$. 
In this section, we plot time in units of $\rho_{0}^{-1/2}$, where
$\rho_{0}$ is the central value of $\rho$ of the corresponding
spherical equilibrium star.

Obviously, models A and B oscillate stably, for both values of $K$. 
The period of these oscillations can be compared with the approximate
analytic value~\cite{compact}
\beq
t_{\rm osc} \simeq 2\pi \Biggl[ {3(\Gamma-1)M^2 \over (5\Gamma-6)RI}
\biggl(3 \Gamma-4-6.75{M \over R}\biggr)\Biggr]^{-1/2},\label{osci}
\eeq
where $M$ is the Newtonian (rest) mass, $R$ is the Newtonian radius of
star, and
\beq
I=\int \rho r^2 dV
\eeq
is the spherical mass moment.  Inserting the values $\rho_c \simeq
5.6M/R^3$ and $I \simeq 0.17MR^2$ for $\Gamma=1.4$, we find
\beq \label{t_osc}
t_{\rm osc} \simeq 5.5 \rho_c^{-1/2} 
	\biggl(0.2 - 6.75{M \over R}\biggr)^{-1/2}.
\eeq
For model A, $M/R \simeq 0.017$ and hence $t_{\rm osc} \simeq 18
\rho_c^{-1/2}$, which is very close to the value that can be read off
in Fig.~2. Obviously, models with a larger compaction $M/R$ have, in
units of $\rho_c^{-1/2}$, a larger oscillation period $t_{\rm osc}$, which
can also be seen in Fig.~2. At the onset of instability, $t_{\rm osc}
\rightarrow \infty$. From equation~(\ref{t_osc}) we therefore find that 
the maximum mass configuration must have a compaction $M/R \simeq 0.029$,
which is very close to that of model C.

Leaving $K=1$ for model C, the star oscillates stably, but reducing
$K$ by only 0.5 \% to 0.995, the central density increases
monotonically, and the star undergoes gravitational collapse.  This
indicates that model C is marginally stable against small
perturbations, $\delta K/K \alt 0.5\%$, and very close to the onset of
instability. This is obviously true, since its proper mass is only
$0.05\%$ smaller than the maximum allowed mass $M_{\rm max}$.

For models D and E, the star monotonically expands or collapses, and
never oscillates: starting with $K=0.995$, the star collapses, and for
$K=1$, the star expands by a large factor\cite{foot2}.  Obviously, for
$K=1$ the star should be in equilibrium, and neither expand nor
collapse. The equilibrium is unstable, however, and even the smallest
truncation error induces a growing perturbation, which must ultimately
lead to gravitational collapse. Initially, this perturbation may be
either an expansion or a contraction.  Since the configuration is
gravitationally bound, the expansion soon has to turn around and lead
to gravitational collapse~\cite{foot2}.  Obviously, we would expect
that for $K>1$ we find an initial expansion, and for $K<1$ an
immediate contraction. However, due to truncation error, the cutoff
between expansion and contraction is not precisely at $K=1$, but
instead at a value slightly smaller than unity ($K \simeq 0.997$ when
we use 30 grid points to cover the star). Again, this value approaches
unity with increasing grid resolution. This behavior establishes that
models D and E are unstable to radial perturbations, which is, of
course, what we expected.

We conclude that we can locate the onset of radial instability, and
in particular that we can determine the maximum allowed mass of neutron
stars to very high accuracy. This will be very important for determining 
the stability properties of binary neutron stars.

\section{Dynamical stability of stars in binary systems}

In this section we present numerical results on the dynamical
stability of stars in binary systems.  We always assume the two stars
to have equal mass, and set up initial data so that they are in a
circular binary orbit. Note that for $\Gamma = 1.4$ the equation of state is
sufficiently soft, so that there is no innermost stable circular
orbit~\cite{lrs93,shibata,BCSSTc}. Therefore, we can choose
arbitrarily small binary separations, and the {\em orbit} of the
binary will still be stable. 
We choose very close binaries, in which the separation of the surfaces 
of the two stars is much smaller than the orbital 
separation ($z_A \simeq 0$ in the terminology of ref.~\cite{BCSSTc}). 
For these binaries the tidal 
effects are strongest, and they are therefore the most suitable
configuration to study the stability against gravitational collapse of
the individual stars.

We evolve three different classes of binary initial data. The first
class are corotational binaries.  For these configurations,
self-consistent equilibrium initial data can be constructed in
post-Newtonian approximation~\cite{shibata} and even in general 
relativity (where the stars are only in quasi-equilibrium,
see~\cite{BCSSTa,BCSSTc}). We denote the class of post-Newtonian
strict equilibrium solutions with a subscript ``a''.

\begin{figure}[t]
\epsfxsize=2.5in
\begin{center}
\leavevmode
\epsffile{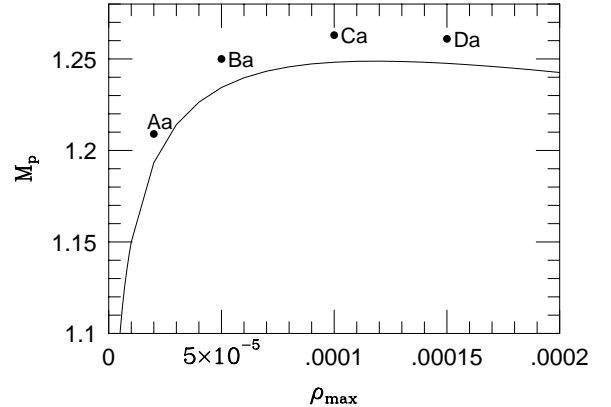}
\end{center}
\caption{Proper mass as a function of the maximum density 
for each star in a corotational equilibrium binary (filled circles) 
and for an isolated spherical star (solid line).}
\end{figure}

In addition to corotational binaries, we would also like to study
irrotational binaries because they are more realistic models for 
binary neutron stars~\cite{cutler}. 
Self-consistent irrotational equilibrium
binaries in Newtonian gravity have recently been constructed by Uryu
and Eriguchi~\cite{UE}, and a relativistic generalization has been
suggested by Bonazzola, Gourgoulhon and Marck~\cite{BGM}. No numerical
models are currently available for such data in a post-Newtonian
approximation. In the absence of self-consistent, tidally deformed
equilibrium data, we therefore take two spherically symmetric stars,
put them close together, and artificially assign an irrotational
velocity profile which maintains their shape and circular orbit
approximately. In order to {\em calibrate} these irrotational initial
data, which are not strictly in equilibrium, we first perform
simulations with a second class of corotational initial data, using
spherically symmetric stars, and compare these with the
self-consistent post-Newtonian equilibrium models which exist in the
corotational case. For this second class, we assign a uniform
velocity
\beq
(u_x, u_y, u_z)=(-\Omega y, \Omega x, 0)
\eeq
to each fluid particle, where $\Omega$ is the orbital
angular velocity. We denote these corotational, near-equilibrium data 
with a subscript ``b''.

Finally, the third class of initial data are irrotational,
near-equilibrium models.  Again, we put two spherically symmetric
stars at a small separation, but now we assign an initial velocity
\beq
(u_x, u_y, u_z)=\biggl(0, \pm \Omega x_0, 0  \biggr)\label{irre}
\eeq
to each fluid particle\cite{foot3}. The centers of mass of the two stars are
located at $(\pm x_0, 0, 0)$, where $x_0 > 0$. The plus sign
in~(\ref{irre}) corresponds to the star at $(+x_0, 0, 0)$, and vice
versa. We denote this third
class of initial data with a subscript ``c''.

In the above velocity profiles, we determine the angular velocity
$\Omega$ from Kepler's law
\begin{equation}
\Omega = \biggl(\frac{M_T}{a^3}\biggr)^{1/2}
\end{equation}
where $M_T=2M_p$ is the sum of the proper mass of the two spherical
stars and $a$ is the coordinate separation between their centers of
mass.

In the above velocity profiles, it would be more physical to fix $v^i$ 
instead of $u_i$. That, however, would involve one more iteration in
the preparation of the initial data, and, in our small compaction 
cases, would make only a negligible difference. We summarize the
initial conditions for six different models, two in each class, in
Table I.

\begin{figure}[t]
\epsfxsize=2.5in
\begin{center}
\leavevmode
\epsffile{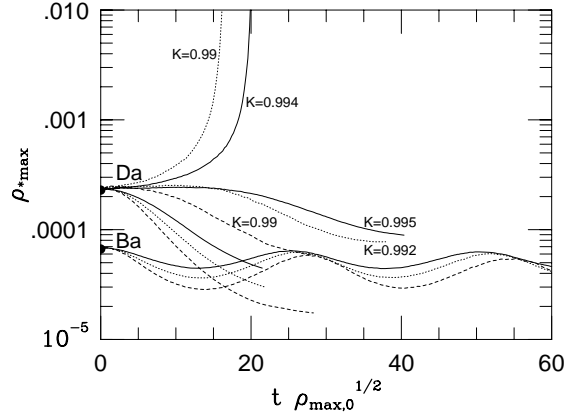}
\end{center}
\caption{Time evolution of $\rho_{* \rm max}$ 
in the corotational equilibrium binary models Ba and Da.  Solid,
dotted and dashed lines denote results obtained with $N=75$, 60, and
50 gridpoints, respectively.  For model Ba, we set $K = 1$
initially. For model Da, initial values of $K$ are 1, 0.995 and 0.994
for $N=75$; 1, 0.992 and 0.99 for $N=60$; and 1 and 0.99 for $N=50$. 
Curves with numerical labels show values of $K \not=1$.
}
\end{figure}

As in our spherical models, we must vary $K$ slightly in order to
investigate the stability of the binary models. Equilibrium stars in
tidal fields are tidally deformed and have a slightly smaller central
density than spherical stars. Using spherical models as initial data
for binaries therefore overestimates the central density, which causes
the stars to expand initially. As we will see, this can be
compensated for by reducing $K$.

\subsection{Corotational equilibrium models}

Following Shibata~\cite{shibata} and Baumgarte {\em
et~al.}~\cite{BCSSTc,BCSSTa}, we construct self-consistent
equilibrium initial data, describing two corotational binary stars in
contact ($z_A=0$). In Fig.~3, we plot the proper mass of each star as a
function of the central density (filled circles), and compare these
values with those for spherical stars in isolation (solid line).  The
maximum allowed mass of binary stars is slightly {\em larger} than
that of spherical stars in isolation (see the discussion
in~\cite{BCSSTc}).

\begin{figure}[t]
\epsfxsize=2.5in
\begin{center}
\leavevmode
\epsffile{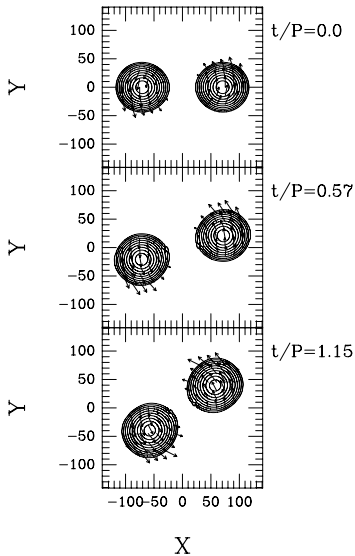}
\end{center}
\caption{Snapshots of density
contour lines and the velocity flow $(v^x, v^y)$ in the equatorial
plane for model Ba.  Contour lines are drawn for
$\rho_*/\rho_{*~0}=10^{-0.3j}$, where $\rho_{*~0}$ denotes the maximum
value at $t = 0$, for $j=0, 1, 2
\cdot\cdot\cdot 10$. Vectors indicate the local velocity field.
Time is shown in units of orbital period ${\rm P}$.  See Table I for
the relation between ${\rm P}$ and $\rho_{{\rm max},0}$.}
\end{figure}

We show results for grid sizes $N=40$, 50, 60, and 75, and find that 
our code is second order accurate. The proper mass $M_p$ evaluated 
on a grid with a grid spacing $d$, $M_p(d)$, therefore scales according
to 
\beq
M_p(d) = M_0 + M_2 d^2 + O(d^3). 
\eeq
Taking values for different $N$ (and hence $d$), we can eliminate the
second order error term by Richardson extrapolating to $d=0$, which
yields the value $M_0$. This is the value that we plotted in
Fig.~3. In Table II we summarize these results by tabulating the
masses $M_p(d)$ for the different grid resolutions, together with the
Richardson extrapolated value $M_0$.

\begin{figure}[t]
\epsfxsize=2.5in
\begin{center}
\leavevmode
\epsffile{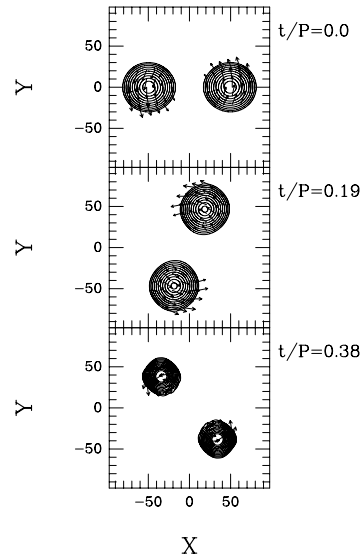}
\end{center}
\caption{Same as Fig.~5, except for model Da at $t=0$. For this
sequence we set $K = 0.994$ initially.}
\end{figure}

Comparing $M_0$ with $M_p(d)$ for $N=60$ and 75, we find deviations of
$\sim 0.6\%$ and $\sim 0.4\%$, respectively. This is a lower limit on
the truncation error that we have to expect in the subsequent
evolution.

Note that in Fig.~3, we plot the mass versus maximum density for a
sequence of constant separation (namely for contact binaries).  In
this graph, the onset of instability need not coincide with the
maximum mass configuration. Instead, the onset of instability can be
located by constructing sequences of constant angular momentum (see
Baumgarte {\em et~al.}~\cite{BCSSTb}). However, we expect that the
onset of instability is very close to the maximum mass
configuration. We therefore present results for two models, one with a
maximum density slightly less than the maximum mass model (denoted Ba,
$\rho_{\rm max} = 5 \times 10^{-5}$), and one with a maximum density
slightly larger (Da, $\rho_{\rm max} = 1.5\times 10^{-4}$).  For both
models, the orbital period is $\sim 50$ in units of 
$\rho_{{\rm max},0}^{-1/2}$ 
where $\rho_{{\rm max},0}$ is maximum value of $\rho$ at $t=0$. 

In Fig. 4, we show the time evolution of the maximum value of
$\rho_{*}$ ($\rho_{* \rm max}$) for models Ba and Da. We show results
for three different grid resolutions, $N=50$, 60, and 75, where we
have kept the location $L$ of the outer boundary constant. We picked
$L$ such that each star is covered by $\sim N/2 -5$
grid points (see Table~I). 

For model Da, we also picked several different values of $K$, and
marked all simulations with $K \not= 1$ accordingly in
Fig.~4. Depending on $K$, these models either collapse or expand, but
never oscillate stably. This indicates that they are dynamically
unstable, as we have expected. As in the discussion of unstable
spherical models, we would expect the cutoff between initial expansion
and contraction to be at $K=1$ if we had arbitrary accuracy. This is
not the case, but we again find that increasing the grid resolution
makes this cutoff approach unity (cutoff value of $K$ 
is less than  $0.99$ for $N=50$, $\simeq 
0.992$ for $N=60$, and $\simeq 0.995$ for $N=75$).

For model Ba, we only show results for $K = 1$. Obviously, this
configuration oscillates stably. The oscillations are due to a slight
inconsistency between the initial data and the evolution scheme. The
amplitude of these oscillations decreases with increasing grid number,
which shows that our method is convergent. Note that $M_0$ of this
configuration is 1.250, which is marginally larger than the maximum
allowed rest mass of an isolated, spherical star, $M_{\rm max} =
1.248$. We therefore conclude that {\em all stars that are stable in
isolation are also stable in a corotational binary}. We expect that the
reverse is not true: a star in a close binary can support more mass
than an isolated, spherical star. Note that our results are different
{}from those of WMM, who found that stars with masses of as much as 10
\% or more below the maximum allowed rest mass in isolation were
destabilized in a close binary. At this level, such an effect would be
discerned easily by our code, but it is not present.

In Figs. 5 and 6 , we show contour lines of $\rho_*$ and the velocity
field of $(v^x, v^y)$ in the equatorial plane for models Ba and Da, 
using a grid resolution of $N=75$. For model Da we set $K=0.994$.
Note that for our adopted soft equation of state, the stars are very
centrally condensed. The tidal field mostly deforms the very low
density envelope, which is hardly visible in these plots. The two
envelopes are pulled towards the binary companion, and in our contact
cases, touch at the origin. The high density cores, however, are
hardly deformed by the tidal field, and are still fairly far
separated.
  
For model Ba, we show contours at $t = 0.0$, $27.0$, and $55.5$, 
all in units of $\rho_{{\rm max},0}^{-1/2}$, which
corresponds to the initial condition, a little more than half an
orbital period, and a little over an orbital period. It is obvious
{}from the graphs that the two stars stably orbit each other. Note also
that the velocity field remains close to being corotational, and that
we do not see any evidence of vorticity features which have been
reported in ref.~\cite{wilson}.

For model Da we show contours at $t = 0.0$ $9.55$
and $19.1\rho_{{\rm max},0}^{-1/2}$, which is a little less than half an
orbital period. It can be seen very clearly how the star contracts and
collapses.

\subsection{Corotational near-equilibrium models}

We now present numerical results for our corotational near-equilibrium
models. We do this to calibrate our code, and to show that our
near-equilibrium models are good approximations to self-consistent
equilibrium models~\cite{foot4}. This justifies studying such
near-equilibrium models for irrotational binaries.

\begin{figure}[t]
\epsfxsize=2.5in
\begin{center}
\leavevmode
\epsffile{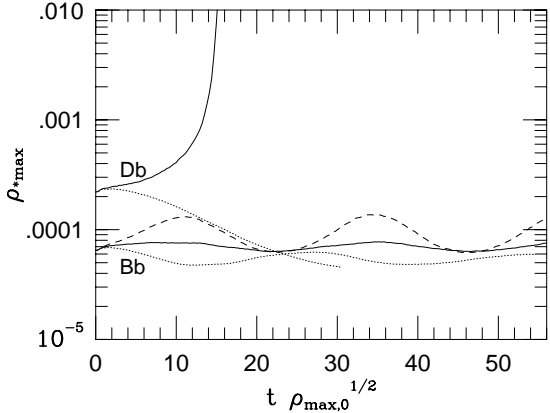}
\end{center}
\caption{Time evolution of  $\rho_{* \rm max}$
in the corotational, near-equilibrium binary
models Bb and Db. Initially, we set $K = 0.98$ (dotted line), 
$K = 0.975$ (solid line) and $K = 0.97$ (dashed line) for model Bb, and 
$K = 0.98$ (dotted line) and $K = 0.975$ (solid line) for model Db. 
For models Bb and Db, the orbital period is 
$\sim 39$ and $46 \rho_{{\rm max},0}^{-1/2}$.}
\end{figure}

\begin{figure}[t]
\epsfxsize=2.5in
\begin{center}
\leavevmode
\epsffile{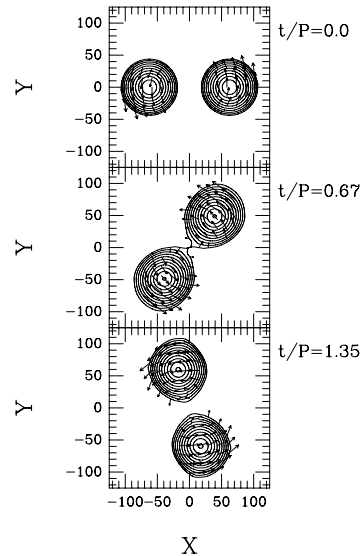}
\end{center}
\caption{Same as Fig.~5, but for model Bb. For this sequence we
set $K=0.975$ initially.}
\end{figure}

For all the simulations discussed in this and the following
subsection, we use a numerical grid with $N=60$ grid points. We adjust
the outer boundary $L$ such that the radius of each star is covered
with $\sim 25$ grid points.  For the models Bb and Bc we used $L=125$ 
and for Db and Dc $L=96$ (see Table~I). We 
construct spherical models, and placed them on the grid such that
their centers of mass is located at $(x,y,z)=(\pm L/2,0,0)$.  Note
that these stars are not in contact. However, the stars are not
tidally deformed, and therefore separation of the center of masses of
the two stars is slightly smaller than for the self-consistent
equilibrium models.  The orbital period for these binaries is $\sim 39$
and 46 in units of $\rho_{{\rm max},0}^{-1/2}$, respectively. 

In Fig.~7, we show the time evolution of $\rho_{* \rm max}$ for models
Bb and Db. Dotted, solid and dashed lines denote results for $K =
0.98$, $0.975$ and $0.97$, respectively. As before, model Db either
expands or contracts, but never oscillates stably. We again conclude that
Db is dynamically unstable. Model Bb, on the other hand, exhibits
stable oscillation for several different values of $K$, and we conclude
that it is dynamically stable.

It is interesting to note that even with a reduced value of $K =
0.98$, model Bb initially expands, albeit stably. Only reducing the
pressure to a value smaller than that ($K \sim 0.975$), the
configuration is roughly in equilibrium.  This can be understood quite
easily, because the spherical star is not a self-consistent
equilibrium solution. The tidal field tends to deform the star, which reduces
the central density. Therefore, our initial data have too high a
central density for their mass, and the star starts expanding. This
can be compensated for by reducing $K$ and hence the pressure. As we
have argued before, reducing $K$ is equivalent to increasing the mass,
and indeed this is another way of understanding why a star in a binary
can support more mass than a star in isolation.

In Fig.~8, we show contour lines of $\rho_*$ and the velocity
field $(v^x, v^y)$ in the equatorial plane for model Bb, where we
have set $K = 0.975$. We show the configuration at $t = 0.0$, 
$26.5$ and $53.0\rho_{{\rm max},0}^{-1/2}$.  Comparing 
these plots with Fig.~5, one can see that the center of masses
of the two stars are closer.  During the evolution, the stars loose
their spherical shape and adjust to the tidal field. However, all the
qualitative features are very similar to the self-consistent 
equilibrium simulations. In particular, the velocity field remains
nearly corotational, and we do not see any evidence of the 
double vorticity fields as reported by WMM.

This test suggest that the spherical near-equilibrium models are very
good approximations to self-consistent equilibrium configurations for
the models and issues we are investigating here (but see
footnote~\cite{foot4}).

\subsection{Irrotational near-equilibrium models}

Viscosities in neutron stars are expected to be too small to bring
neutron star binaries into corotation on the timescale of their
evolution~\cite{cutler}. Numerical models exist only in Newtonian
gravity~\cite{UE,BGM} or in the Newtonian and
post-Newtonian ellipsoidal approximation~\cite{lrs94,lrs97}.  We
therefore adopt the spherical near-equilibrium approximation to
construct initial models, which we have calibrated and found to be
adequate in section~IV.B for corotational cases.

\begin{figure}[t]
\epsfxsize=2.5in
\begin{center}
\leavevmode
\epsffile{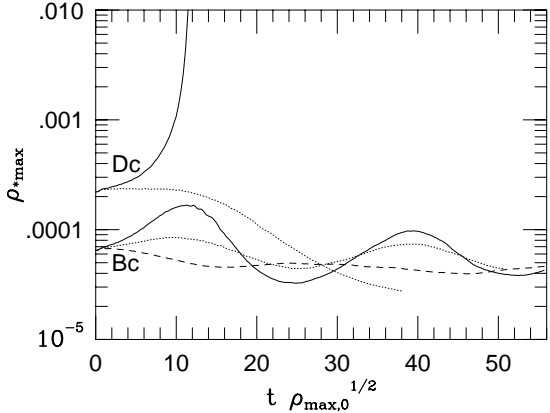}
\end{center}
\caption{ Time evolution of $\rho_{* \rm max}$
in the irrotational, near-equilibrium binary models Bc and Dc.
Initial values of $K$ are $0.99$, $0.985$ and 0.98 for model Bc, and
$0.98$ and $0.985$ for model Dc.  Solid, dotted and dashed lines
denote results for $K=0.98$, 0.985 and 0.99, respectively. The orbital
period of models Bc and Dc is $\sim 39$, and 
$46\rho_{{\rm max},0}^{-1/2}$, again.}
\end{figure}

\begin{figure}[t]
\epsfxsize=2.5in
\begin{center}
\leavevmode
\epsffile{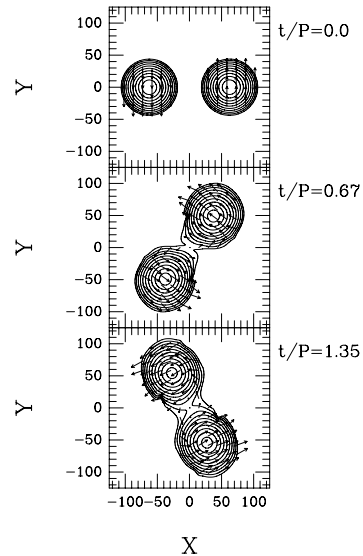}
\end{center}
\caption{Same as Fig.~5, but for model Bc. For this sequence, we set 
$K=0.99$ initially.}
\end{figure}

We again vary $K$ and choose $K=0.99$, $0.985$ and $0.98$ for model
Bc, and $K=0.985$ and $0.98$ for model Dc.  In Fig. 9, we show time
evolution of $\rho_{* \rm max}$ for models Bc and Dc.  Dashed, dotted
and solid lines denote results for $K = 0.99$, $0.985$ and $0.98$,
respectively.  As in the corotational cases, model Dc cannot be held
stably, whereas model Bc oscillates for all these choices of $K$.
Therefore, we again conclude that model Bc is dynamically stable,
whereas Dc is not.

It is interesting to note that for these irrotational models, we had
to reduce $K$ by a somewhat smaller amount to minimize the amplitude
of oscillations in model Bc than for the corotational model Bb ($K
\sim 0.985$ here, and $K \sim 0.975$ for model Bb). This can be
understood very easily, because in corotational models, the individual
stars are spinning, and are therefore stabilized and deformed by both
the tidal field and the their own spin. In irrotational binaries, the
stars have almost no spin (with respect to distant inertial
observers), and are deformed only by the tidal field. Therefore,
putting a star into an irrotational binary will reduce the central
density by less than putting the same star into a corotational
binary. As a consequence, we have to reduce $K$ by a smaller amount to
compensate.  Applying our scaling argument, this result means that a
irrotational binary can support less mass than a corotational binary,
but still more than a spherical star in isolation.  This result is
corrobated by the post-Newtonian ellipsoidal models constructed
in~\cite{lrs97}.

In Fig. 10, we show contour lines of $\rho_*$ and the velocity
field $(v^x, v^y)$ in the equatorial plane for model Bc and $K=0.99$
initially. We show contours at $t = 0.0$, $26.5$, and 
$53.0\rho_{{\rm max},0}^{-1/2}$. For the bulk of the matter at 
the core of the stars, the velocity field remains approximately
irrotational, and the stars stably orbit each other. We conclude
that there is no qualitative difference between corotational and
irrotational binaries as far as their radial stability properties are 
concerned.

\section{Summary}

We perform post-Newtonian, dynamical simulations of close binaries in
circular orbit. In particular, we study the stability of the
individual stars against gravitational collapse in both corotational 
and irrotational systems containing stars of equal mass.

We have chosen a soft, adiabatic equation of state with $\Gamma =
1.4$, for which there is no innermost stable circular orbit, so that
the binary {\em orbit} is stable even when the stars are in contact,
and for which the onset of instability for a spherical star in
isolation occurs at a very small value of the compaction $M/R$. We can
therefore study the individual stars' stability properties in near
contact binaries, for which the tidal effects are strongest, and in a
regime in which a post-Newtonian approximation is very accurate.

We do not find any crushing effect as reported by WMM~\cite{wilson}.
In contrast, the maximum density in both corotational and irrotational
binaries is smaller than that of spherical stars in isolation.  We
find that stars in binaries can support more mass than in isolation.
Moreover, all stars that are stable against radial perturbations in
isolation, will also be dynamically stable when put into a binary.

All these results are in complete agreement with, for example, the
findings of Baumgarte {\em et~al.}~\cite{BCSSTa,BCSSTb},
Flanagan~\cite{EEF}, and Thorne~\cite{KIP}.  For the most part, their
discussions rigorously address {\em secular} stability only. Several
different arguments can be invoked to suggest that {\em secularly}
stable binaries are also {\em dynamically} stable, but this is
strictly proven only in Newtonian theory (see also~\cite{lrs97}).  Our
dynamical calculations reported in this paper are the first to
directly confirm dynamical stability, at least within our
post-Newtonian approximation.

We compare, in a near-equilibrium approximation, corotational and
irrotational binary models.  As expected, stars in corotational
binaries can support slightly more mass than in irrotational binaries,
but apart from these small differences we do not find any qualitative
difference in their radial stability properties.  A more rigorous
treatment will require the construction of post-Newtonian, 
irrotational equilibrium binary models for initial
data.

Since our computations have been performed in nondimensional units,
our results apply not only to neutron star binaries, but also to
binaries of white dwarfs and supermassive stars. In fact, the
equations of state of massive white dwarfs (ideal degenerate,
extremely relativistic electrons) and supermassive stars (radiation
$\gg$ thermal pressure) are closely approximated by the value $\Gamma
= 1.4$ that we have adopted.  These binaries may be important low
frequency gravitational wave sources for future space-based
gravitational waved detectors, like LISA.

\acknowledgments

Numerical computations were performed on the FACOM VX/4R machine in
the data processing center of the National Astronomical Observatory of
Japan.  M.~S. wishes to express his gratitude to the University of
Illinois at Urbana-Champaign for its hospitality during his stay in
September, 1997, when this project was initiated.  This work was
supported by a Japanese Grant-in-Aid of the Ministry of Education,
Culture, Science and Sports (Nos.~08NP0801 and 09740336), and by NSF
Grant AST 96-18524 and NASA Grant NAG 5-3420 at Illinois.

\twocolumn[\hsize\textwidth\columnwidth\hsize\csname
@twocolumnfalse\endcsname

\vskip 5mm
\noindent
{\bf Table I.~} Initial conditions of binary models. We tabulate the
maximum density $\rho_{{\rm max},0}/10^{-5}$, the radius $R_{\infty}$
of a spherical star of the same rest mass in isolation, the orbital
period ${\rm P}$ in units of $\rho_{{\rm max},0}^{-1/2}$, the nature
of the initial velocity field and matter profile, and the location of
the outer boundary $L$ (see text). All quantities are shown in units
of $c=G=K=1$.

\vskip 5mm
\begin{center}
\begin{tabular}{|c|c|c|c|c|c|c|} \hline
~Model~ & ~$\rho_{{\rm max},0}/10^{-5}$~ & $~~R_{\infty}~~$ & 
 ${\rm P}\times \sqrt{\rho_{{\rm max},0}}$ 
& ~ Velocity Field~ & Matter Profile 
	& $~~L~~$\\ \hline
Ba    &$5 $ &53 &48.6& corotation   & equilibrium &~ 141~\\ \hline 
Da    &$15$ &38 &50.2& corotation   & equilibrium & 97.5 \\ \hline
Bb    &$5$  &53 &39.4& corotation   & spherical & 125 \\ \hline
Db    &$15$ &38 &45.9& corotation   & spherical & 96 \\ \hline
Bc    &$5$  &53 &39.4& irrotation   & spherical & 125 \\ \hline
Dc    &$15$ &38 &45.9& irrotation   & spherical & 96 \\ \hline
\end{tabular}
\end{center}

\vskip 5mm
\noindent
{\bf Table II.~} Proper mass $M_p$ of each star in a corotational
equilibrium binary for different grid resolutions. We tabulate
$M_p$ versus central density $\rho_c$ for 
$N=40$, 50, 60, and 75. We also list the Richardson extrapolated
value $M_0$ as well as the proper mass of the spherical model
in isolation with the same central density.

\vskip 5mm
\begin{center}
\begin{tabular}{|c|c|c|c|c|} \hline
$\rho_c/10^{-5}$ & $2 $ & $5 $ &
$10$ & $15$    \\ \hline
~~$N=40$~~ &~~ 1.1954~~&~~ 1.2347~~ &~~ 1.2469~~
&~~ 1.2451~~ \\ \hline
~~$N=50$~~ &~~ 1.2004~~&~~ 1.2402~~ &~~ 1.2526~~
&~~ 1.2509~~ \\ \hline
~~$N=60$~~ &~~ 1.2031~~&~~ 1.2431~~ &~~ 1.2556~~
&~~ 1.2539~~ \\ \hline
~~$N=75$~~ &~~ 1.2051~~&~~ 1.2455~~ &~~ 1.2581~~ 
&~~ 1.2565~~ \\ \hline
$M_0$                    & 1.209  & 1.250  & 1.263  & 1.261  \\ \hline
Spherical &~~ 1.1933~~ &~~ 1.2344~~ &~~ 1.2482~~ &~~ 1.2476 ~~\\ \hline
\end{tabular}
\end{center}

\vskip2pc]


\begin{thebibliography}{99}
\bibitem{foot0}
Newtonian and post-Newtonian considerations illustrate the physical 
meaning of $_{(n)}\alpha$, $_{(n)}\psi$, $_{(n)}\beta^i$, and 
$_{(n)}h_{ij}$ for $n\not=0$.  The terms $_{(2)}\alpha$, $_{(2)}\psi$, 
$_{(1)}\beta^i$, and $_{(2)}h_{ij}$ do not affect the dynamics of the stars; 
effects due to tidal deformation of spherical stars appear at $n=6$ 
in $_{(n)}\alpha$ and $_{(n)}\psi$, at $n=7$ in $_{(n)}\beta^i$, and 
at $n=8$ in $_{(n)}h_{ij}$; effects of gravitational radiation 
are included in $_{(n)}h_{ij}$ for $n \geq 5$, 
in $_{(n)}\alpha$ and $_{(n)}\psi$ for $n \geq 7$, 
and in $_{(n)}\beta^i$ for $n \geq 8$~\cite{ASF}.  We also note that in 
Eqs.~(\ref{MAA})$-$(\ref{MAE}), gauge dependent terms (e.g., terms 
such as $_{(5)}h_{ij}$ that may be erased by a gauge transformation) 
are included, because they appear in our gauge condition.
\bibitem{foot2}
Note that for the case where the star expands, we stopped the
computation when the surface of the star reached the outer boundary of
numerical grid. Since the star is gravitationally bound, 
the star ultimately has to recontract and collapse.
\bibitem{foot4}
For stiffer equations of state with higher $\Gamma$, the bulk of the mass
would suffer greater tidal distortion at the same separation. Our
near-equilibrium spherical approximation would require larger
orbital separations, and hence more numerical resolution, to be
as reliable in these cases.
\end{thebibliography}
\end{document}